\documentclass[prl,aps,floatfix,showpacs,superscriptaddress,twocolumn]{revtex4}
\usepackage{graphicx,dcolumn,url}
\def \Pt {{P}_{T}}
\def \Et {{E}_{T}}
\def \met  {\,/\!\!\!\!E_{T}}
\begin{document}                                             
 
\title{
% \begin{flushright}
% {\small /CDF/PUB/EXOTIC/CDFR/7374 \\ PRL draft v3 \\ \today\\ }
% \end{flushright}
Search for Higgs Bosons Decaying into $b\bar{b}$ and Produced in Association 
with a Vector Boson in $p\bar{p}$ Collisions at $\sqrt{s} = 1.8$~TeV
}
\author{D.~Acosta}
\affiliation{University of Florida, Gainesville, Florida  32611}
\author{T.~Affolder}
\affiliation{University of California at Santa Barbara, Santa Barbara, 
California 93106}
\author{M.G.~Albrow}
\affiliation{Fermi National Accelerator Laboratory, Batavia, Illinois 60510}
\author{D.~Ambrose}
\affiliation{University of Pennsylvania, Philadelphia, Pennsylvania 19104}
\author{D.~Amidei}
\affiliation{University of Michigan, Ann Arbor, Michigan 48109}
\author{K.~Anikeev}
\affiliation{Massachusetts Institute of Technology, Cambridge, Massachusetts  
02139}
\author{J.~Antos}
\affiliation{Institute of Physics, Academia Sinica, Taipei, Taiwan 11529, 
Republic of China}
\author{G.~Apollinari}
\affiliation{Fermi National Accelerator Laboratory, Batavia, Illinois 60510}
\author{T.~Arisawa}
\affiliation{Waseda University, Tokyo 169, Japan}
\author{A.~Artikov}
\affiliation{Joint Institute for Nuclear Research, RU-141980 Dubna, Russia}
% ...
\collaboration{CDF Collaboration}
\noaffiliation

\date{\today}

\begin{abstract}
We present a new search for $H^{0}V$ production, where $H^{0}$ is a scalar Higgs 
boson decaying into $b\bar{b}$ with branching ratio $\beta$, and $V$ is a 
$Z^{0}$ boson decaying into $e^{+}e^{-}$, $\mu^{+}\mu^{-}$, or $\nu\bar{\nu}$.  
This search is then combined with previous searches for $H^{0}V$ where $V$ is a 
$W^{\pm}$ boson or a hadronically decaying $Z^{0}$.  The data sample consists of $106 \pm 4$ 
pb$^{-1}$ of $p\bar{p}$ collisions at $\sqrt{s}=1.8$ TeV accumulated by the 
Collider Detector at Fermilab.  Observing no evidence of a signal, we set 95\% 
Bayesian credibility level upper limits on $\sigma(p\bar{p}\rightarrow H^{0}V)
\times\beta$.  For $H^0$ masses of 90, 110 and 130 GeV/$c^{2}$, the limits are 
7.8, 7.2, and 6.6\,pb respectively.
\end{abstract}                           

\maketitle

% introduction
% ------------
A key component of the standard model (SM) is spontaneous electroweak
symmetry breaking, which gives rise to the mass of all fermions and the 
$W^\pm$ and $Z^0$ gauge bosons. This process leads to the existence of a 
neutral scalar particle, the Higgs boson ($H^0$), whose mass is unspecified in 
the SM, but whose couplings to all other particles of known mass are fully 
specified at tree level. The Higgs boson has not been directly observed, but 
its expected contribution to loop corrections for many SM observables has allowed
an inferred mass of $M_{H} = 126^{+73}_{-48}$ GeV/$c^{2}$ from precision 
electroweak measurements~\cite{ewprec}. In addition, direct searches at LEP\,2 
have excluded, with a 95\% confidence level, a SM Higgs boson with 
$M_{H} < 114.4$ GeV/$c^{2}$~\cite{lep2lim}.  The relatively low $H^0$ mass 
favored within the SM framework implies the possibility of its direct 
observation at the Tevatron in Run\,II, where searches have begun by both the 
D\O~\cite{d0run2} and CDF collaborations.  Here we report on direct searches 
using data accumulated by the Collider Detector at Fermilab (CDF) between 
February 1992 and July 1995 (Run I) for a total integrated luminosity of 
$106 \pm 4$ pb$^{-1}$. 

At the Tevatron, the SM Higgs boson is produced from both gluon-gluon and
quark-antiquark initial states~\cite{stange}. Although the dominant production
mechanism is $gg \rightarrow H^0$, production in association with vector
bosons ($q\bar{q}^\prime \rightarrow H^0 W^\pm$, $q\bar{q} \rightarrow
H^0 Z^0$) provides the most sensitive channels for Higgs boson searches at the
Tevatron if $M_{H} < 140$ GeV/$c^{2}$, because one can obtain significant 
background rejection from the additional highly energetic objects in the event 
coming from the vector boson decays. The predicted cross section, $\sigma_{V\!H^0}$, for 
$VH^0$ production from $p\bar{p}$ collisions at $\sqrt{s}=1.8$\,TeV varies 
between 0.50 and 0.15\,pb for $H^0$ masses between 90 and 130\,${\rm GeV}/c^2$, 
with the ratio $\sigma_{W^\pm H^0} / \sigma_{Z^0 H^0} \approx 1.6$.

% Our search strategy is to look for $VH^0$ production with 
% $H^0 \rightarrow b\bar{b}$ and vector boson decays in either the hadronic
% or leptonic channels. The $W^\pm$ and $Z^0$ decay products define the various
% search channels. For $90 < M_H < 130\,{\rm GeV}/c^2$ the branching ratio
% ($\beta$) for $H^0 \rightarrow b\bar{b}$ varies between 0.85 and 0.60~\cite{hbr}.
 
We have previously reported the results of searches in the 
$WH^{0}\rightarrow\ell\nu\,b\bar{b}$ ($\ell = e$ or $\mu$) and $VH^{0}\rightarrow 
q\bar{q}^{\prime}\,b\bar{b}$ channels~\cite{vhlnbb,vhqqbb}. Here 
we add the searches for $Z^{0}H^{0}$ production using the decay channels
$\ell^{+}\ell^{-}\,b\bar{b}$  and $\nu\bar{\nu}\,b\bar{b}$.
Finding no evidence for Higgs boson production using these decay modes, we set 
limits on the production cross section as a function of mass, and combine our 
results with the previous $VH^0$ cross section limits. These limits represent
the final CDF cross section limits for Higgs boson production in association 
with a vector boson from the Run I data.

% detector
% --------
The CDF detector is described in Ref.\cite{detector}, and the coordinate system 
and various quantities used throughout this paper are defined in Ref.~\cite{eta}.
The momenta of the charged leptons are measured with the 
central tracking chamber in a 1.4\,T superconducting solenoidal magnet.  
Electromagnetic and hadronic calorimeters surrounding the tracking chambers 
are used to identify electrons and jets and measure their energies.  
Muons are identified with drift chambers located outside the calorimeters.  
The silicon vertex detector (SVX)
%, consisting of four layers of axial
%microstrips at radii ranging from 2.9\,cm to 10.5\,cm, 
is the innermost
detector used for precise tracking in the plane transverse to the 
beam~\cite{svx}.

In the analyses reported here, two algorithms using tracks measured
with the SVX are applied
to identify jets originating from heavy flavor quarks ($b$ and $c$).
The first reconstructs a secondary
vertex (a vertex displaced from the primary interaction vertex) produced 
by the heavy flavor decays and measures the transverse decay length (SVX tag). 
The resolution of the transverse decay length of the secondary vertex is 
typically of order $150\,\mu$m.  
The second algorithm 
uses the impact parameter of the tracks in the jet (the closest 
distance of the track to the primary vertex in the transverse plane) 
to calculate a probability that the jet is not from heavy flavor (JPB tag).
The details of these tagging algorithms are given in Ref.~\cite{topxs2}. 

%
%Tracks measured with the SVX are used in the reconstruction of secondary
%vertices (vertices displaced from the primary interaction vertex) produced 
%by heavy flavor ($b$ and $c$) meson decays. The resolution of
%the transverse decay length of the secondary vertex is typically of
%order $150\,\mu$m.  
%In the analyses reported here, jets are tagged as originating from 
%heavy flavor quarks by two algorithms. The first reconstructs a secondary vertex 
%and measures the transverse decay length (SVX tag). The second  
%uses the impact parameter of the tracks in the jet (the closest 
%distance of the track to the primary vertex in the transverse plane) 
%to calculate a probability that the jet is not from heavy flavor (JPB tag).
%%or identifying a soft lepton in the jet (SLT tag). 
%The details of these tagging algorithms are given in Ref.~\cite{topxs2}. 
%

% event selection - llbb
% ----------------------
For the details of the analyses previously published we refer
to those publications~\cite{vhlnbb,vhqqbb}, and list the results here
when appropriate, as they are used in the combined cross section limits. 
We now describe the two new channels, $\ell^{+}\ell^{-}\,b\bar{b}$
and $\nu \bar{\nu}\,b\bar{b}$.
Events for the $\ell^{+}\ell^{-}\,b\bar{b}$ channel analysis are required to
pass a high-$\Pt$ lepton trigger and must contain two high-$\Pt$~\cite{eta}, 
oppositely charged leptons ($e$ or $\mu$) that are isolated from nearby tracks 
and calorimeter activity.
%Event selection for the $\ell^{+}\ell^{-}\,b\bar{b}$ channel analysis begins
%with two high-$\Pt$~\cite{eta}, oppositely charged leptons ($e$ or $\mu$), that
%must be isolated from nearby tracks and calorimeter activity.
At least one lepton is required to have $\Pt > 20$ GeV/$c$ 
and be in the central detector ($|\eta|< 1.0$).  For the second 
lepton the $\Pt$ requirement is relaxed to 10 GeV/$c$ and the pseudorapidity 
range is extended into the plug calorimeter, up to $|\eta|\sim 2.4$.  
The dilepton invariant mass must be in the range $76 < M_{\ell\ell} < 106$\,GeV/$c^2$
to be consistent with the decay of a $Z^0$ boson. This requirement essentially
removes any sensitivity of this analysis to $Z^0 \rightarrow \tau^+ \tau^-$.
The event is additionally required to contain two or three high-$\Et$ jets 
($\Et > 15$ GeV), at least one of which is SVX tagged. A cut on the missing 
transverse energy ($\met < 50$ GeV)~\cite{eta} is also applied, with the effect of
reducing the $t\bar{t}$ background by approximately a factor of two, while 
preserving about 95\% of the signal. 

The $\nu \bar{\nu}\,b\bar{b}$ channel is characterized by two heavy flavor
jets and large $\met$ from the neutrinos. The data sample for this channel
is derived from an event trigger requiring $\met > 35\,{\rm GeV}$
in addition to event quality cuts~\cite{metjet}.
To reject $W^\pm$ and $Z^0$ decays to leptons, events containing an isolated 
track with $\Pt > 10\,{\rm GeV}/c$ are removed from the sample.
To ensure less susceptibility to the uncertainty in the trigger efficiency
at threshold, the analysis requires $\met > 40\,{\rm GeV}$. The
trigger efficiency is approximately 60\% at this value of $\met$.  
Additionally the event must contain two or three jets with 
$\Et > 15\,{\rm GeV}$ (about 20\% of the $ZH^0$ signal contains a third jet). 
To reject QCD multi-jet events where the $\met$
results from a mismeasured jet, the azimuthal angle between the $\met$
and the direction of any jet with $\Et > 8\,{\rm GeV}$ is required to be
at least 1.0 radians. In addition, the jets from inclusive di-jet production tend 
to be back-to-back, while jets from $H^0 \rightarrow b\bar{b}$ in $ZH^0$ events
tend to have a smaller opening angle, leading to the requirement that the 
azimuthal angle between the leading two jets be less than 2.6 radians.
Approximately 10\% of the efficiency from the $\nu\bar{\nu}\,b\bar{b}$ selection 
is contributed by $W^\pm H^0$ events where the lepton is undetected.
   
Events in the $\nu \bar{\nu}\,b\bar{b}$ sample are classified as 
``single-tagged'' (exactly one SVX tagged jet) or ``double-tagged'' (one SVX 
tagged jet and a second jet tagged by either the SVX or JPB tagging algorithms). 
The backgrounds and efficiencies are calculated separately for these orthogonal 
sets, which are then treated as separate but correlated channels when combined 
with the other channels. This is analogous to what was done in the 
$WH^{0}\rightarrow\ell\nu\,b\bar{b}$ search~\cite{vhlnbb}.
%
%efficiency and acceptance
%-------------------------
%
\begin{table}[htb]
\caption{\label{table:vhacc}
Total selection efficiencies for $VH^0$ events in each analysis channel used 
in the combined result, as a function of the $H^0$ mass, $M_H\,({\rm GeV}/c^2)$. 
Numbers are percentages and include the branching ratios of the vector boson 
($W^\pm$ or $Z^0$) in a given channel.  ST refers to single-tagged events and 
DT to double-tagged events.  Uncertainties include systematic effects.}
\begin{ruledtabular}
\begin{tabular}{l|ccc}
         & \multicolumn{3}{c}{$VH^0$ event efficiencies (\%)}  \\
Channel  &  $M_H = 90$    &  $M_H = 110$   & $M_H = 130$       \\ 
\hline
$\ell^+\ell^-\,b\bar{b}$ 
         & $0.14\pm 0.03$ & $0.20\pm 0.04$ & $0.19\pm 0.04$    \\
$\nu\bar{\nu}\,b\bar{b}$ (ST)
         & $0.51\pm 0.10$ & $0.63\pm 0.13$ & $0.76\pm 0.15$    \\
$\nu\bar{\nu}\,b\bar{b}$ (DT)
         & $0.37\pm 0.08$ & $0.43\pm 0.09$ & $0.51\pm 0.10$    \\
$\ell\nu\,b\bar{b}$ (ST)
         & $0.59\pm 0.15$ & $0.72\pm 0.18$ & $0.80\pm 0.20$    \\
$\ell\nu\,b\bar{b}$ (DT)
         & $0.22\pm 0.06$ & $0.29\pm 0.07$ & $0.30\pm 0.08$    \\
$q\bar{q}^{\prime}\,b\bar{b}$
         & $1.3\pm 0.7$   & $2.2\pm 1.1$   & $3.1\pm 1.6$      \\
\end{tabular}
\end{ruledtabular}
\end{table}

The efficiencies for identifying $VH^0$ events with our selection criteria 
are summarized in Table~\ref{table:vhacc} and are determined from a 
{\sc pythia}~\cite{Pythia} Monte Carlo simulation of Higgs boson production 
via $V^{\ast}\rightarrow VH^{0}\rightarrow Vb\bar{b}$ followed by a detector 
simulation.  The Higgs boson is forced to decay to $b\bar{b}$ with a 100\% 
branching ratio.  The identification efficiencies for single leptons are 
measured from $Z^0 \rightarrow \ell^{+}\ell^{-}$ events in the data and are 
found to be $91$\% for muons and $83$\% for electrons~\cite{dilprl}.
The SVX and JPB $b$-tagging efficiencies are determined using
data and Monte Carlo samples with high $b$-purity~\cite{topxs2}. 
In the $\ell^{+}\ell^{-}\,b\bar{b}$ channel, the efficiency for obtaining
$\ge 1$ SVX tag in a signal event is $(45 \pm 7)$\%.
The double $b$-tagging efficiency in the $\nu\bar{\nu}\,b\bar{b}$ channel
(SVX\,+\,SVX or SVX\,+\,JPB) is $(19 \pm 4)$\%, and the single $b$-tagging
efficiency (one SVX tag) is $(25 \pm 3)$\%.
The total event efficiencies are the product of the trigger efficiencies, the 
kinematic and geometric acceptances from the selection cuts, the lepton 
identification efficiencies when appropriate, the $b$-tagging efficiencies,
and the $V$ branching ratio relevant for a given search channel.
The systematic uncertainties in the total efficiencies for the 
$\ell^+\ell^-\,b\bar{b}$ and $\nu\bar{\nu}\,b\bar{b}$ channels 
are approximately $20\%$, comprised mostly of the uncertainties in the 
$b$-tagging efficiency (15\%), the modeling of initial and final state
radiation (7\%), lepton identification efficiency (7\% for the 
$\ell^+\ell^-\,b\bar{b}$ channel), trigger efficiency (5\% for the
$\nu\bar{\nu}\,b\bar{b}$ channel), integrated luminosity (4\%) and
the energy scale of jets (3\%).

% backgrounds
%------------   
In the $\ell^+\ell^- b\bar b$ channel the dominant background is
$Z^0$ production in association with a heavy flavor pair 
($Z^0 b\bar{b}$, $Z^0 c\bar{c}$), which accounts for approximately 60\% 
of the total. About 20\% comes from $Z^0 + $jets events where a jet is 
mistagged due to track mismeasurements, and there are smaller contributions 
from $Z^0 c$, $Z^0 b$, diboson, and $t\bar{t}$. All backgrounds are
determined using Monte Carlo simulations except that from $Z^0 + $jets
which uses the data.
%
%backgrounds result predominantly from $Z^0$ production in association with 
%heavy quarks ($Zb\bar{b}$, $Zc\bar{c}$, $Zc$), $V + $jets events where a 
%jet is mistagged due to track mismeasurements, and smaller contributions from
%diboson production, $t\bar{t}$ and single top production.
%%%%

The $\nu\nu b\bar b$ channel background is dominated by QCD jet production of 
$b\bar b$ where the $\met$ results from mismeasured jets.  To calculate this 
contribution we first parameterize the tagging rate in the $\met < 40$ GeV 
region of the data as a function of jet $\Et$ and track multiplicity.  By applying this
parametrization to the jets in the signal region we estimate the QCD
background to be about 70\% of the total background in the single-tagged sample 
and about 50\% in the double-tagged sample. Smaller backgrounds
include $V +$ heavy flavor, diboson, and $t\bar{t}$, all of which are derived
from Monte Carlo simulations.
%%%%

\begin{table}[htb]
\caption{\label{table:bckgrd}
Predicted numbers of events in each channel from all backgrounds 
(see text),
% which includes $W/Z\,b\bar{b}$, $W/Z\,c\bar{c}$, $W/Z\,c$, mistagged
% events, diboson production ($WZ$/$ZZ$/$WW$), and top ($t\bar{t}$ and
% single top production)), 
expected number of signal events for $M_H = 110$ GeV/c$^2$, and number of events 
observed.  Uncertainties include systematic effects.  There is no reliable 
prediction for the background in the $q\bar{q}^{\prime}\,b\bar{b}$ channel.}
\begin{ruledtabular}
\begin{tabular}{l|D{,}{}{-1}cD{.}{}{-1}}
Channel     & \text{Backgr},\text{ound} & Signal   & \text{Dat}.\text{a} \\ 
\hline
$\ell^+\ell^-\,b\bar{b}$       & 3.2\:\pm,\: 0.7 & $0.06\pm 0.01$ &   5. \\
$\nu\bar{\nu}\,b\bar{b}$ (ST)  & 39\:\pm,\: 4    & $0.20\pm 0.04$ &  40. \\
$\nu\bar{\nu}\,b\bar{b}$ (DT)  & 3.9\:\pm,\: 0.6 & $0.14\pm 0.03$ &   4. \\
$\ell\nu\,b\bar{b}$ (ST)       & 30\:\pm,\: 5    & $0.23\pm 0.06$ &  36. \\
$\ell\nu\,b\bar{b}$ (DT)       & 3.0\:\pm,\: 0.6 & $0.09\pm 0.02$ &   6. \\
$q\bar{q}^{\prime}\,b\bar{b}$  &                 & $0.73\pm 0.29$ & 589. \\
\end{tabular}
\end{ruledtabular}
\end{table}
For each decay channel, Table~\ref{table:bckgrd} summarizes the total expected 
backgrounds, the expectations from standard model $VH^0$ production for 
$M_H = 110$ GeV/$c^2$ and $H^0 \rightarrow b\bar{b}$, and the number of data
events observed.  The dominant background in the $q\bar{q}^{\prime}b\bar{b}$ 
channel is QCD production of $b\bar{b}$ with additional jets, hereafter 
abbreviated as ``QCD''. Its normalization is difficult to predict and therefore 
left unconstrained in the analysis.
%where the predominant background 
%is due to mistags and difficult to measure, the total background is normalized 
%from the data. 
%The effect of the new background evaluations reported in~\cite{topxs2}
%are not included here.  The overall effect was to increase the backgrounds
%in the single tag channels by about 10\% and those in the double tag
%channels by about 20\%. This is within the reported background
%uncertainties and the limits obtained are relatively insensitive to this
%difference in background estimates. 
Further details of the background calculations are given 
elsewhere~\cite{vhlnbb,vhqqbb,topxs2}.

% results
%--------

\begin{figure}[htb!]
\includegraphics[width=\columnwidth]{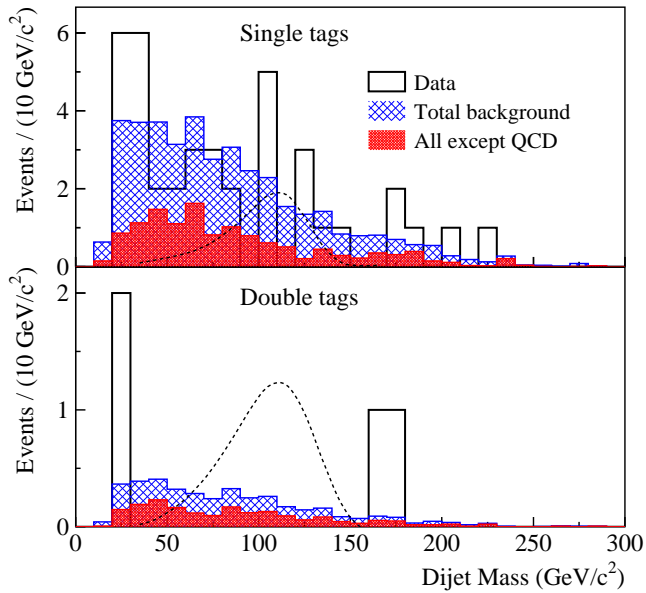}
\caption{\label{fig:vh_fig1}
Dijet invariant mass in $\nu \bar{\nu}\,b\bar{b}$ candidate events,
for events with exactly one $b$-tagged jet and separately for events
with two $b$-tagged jets. The single $b$-tag data includes one overflow 
event. The background shapes shown differ only in the inclusion of the 
predominant background of QCD $b\bar{b}$ production. The signal shape 
shown (dashed line) is for a SM Higgs mass of $110\,{\rm GeV}/c^2$ and a
normalization of 50 times the expected rate.}
\end{figure}

\begin{figure}[htb!]
\includegraphics[width=\columnwidth]{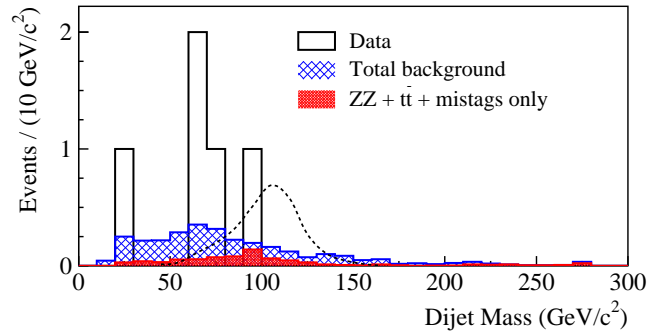}
\caption{\label{fig:vh_fig2}
Dijet invariant mass in $\ell^+ \ell^-\,b\bar{b}$ candidate events.
At least one jet is required to be $b$-tagged by the SVX algorithm. 
The background shapes shown differ only in the inclusion of the 
predominant background of $Z^0+$heavy flavor. The signal shape 
shown (dashed line) is for a SM Higgs mass of $110\,{\rm GeV}/c^2$ and a
normalization of 50 times the expected rate.}
\end{figure}

% We set upper limits on the production cross section of $VH^0$ times the
% branching ratio of $H^0 \rightarrow b\bar{b}$ for each of the $V$ decay 
% channels as a function of the $H^0$ mass.  
A binned likelihood is used to compare the dijet mass spectrum (of the two 
tagged jets, or the one tagged jet and highest-$\Et$ untagged jet) in the 
data to a combination of expected distributions from the background processes 
and the $VH^0$ signal, as a function of $H^{0}$ mass.  The observed dijet mass 
spectra for the $\nu\bar{\nu}\,b\bar{b}$ and $\ell^{+}\ell^{-}\,b\bar{b}$ 
channels are shown together with the expected background shapes in 
Figs.~\ref{fig:vh_fig1} and~\ref{fig:vh_fig2} respectively.

Since no signal is observed, we calculate upper limits on $VH^{0}$
production using a Bayesian procedure.  For each channel, a posterior
density is obtained by multiplying the likelihood function for that
channel with prior densities for all the parameters in the likelihood: 
integrated luminosity, background normalizations, signal efficiency, and 
the product $\sigma_{V\!H^{0}}^{\prime}\equiv\sigma_{V\!H^{0}}\times\beta$ 
of the signal cross section $\sigma_{V\!H^{0}}$ by the branching ratio $\beta$ for 
$H^{0}\rightarrow b\bar{b}$.  With two exceptions, these priors are truncated 
Gaussian densities constraining a given parameter to its expected value within its 
uncertainty.  The exceptions are $\sigma_{V\!H^{0}}^{\prime}$ and the QCD background 
normalization in the $q\bar{q}^{\prime}\,b\bar{b}$ channel.  Since nothing is 
presumed known a priori about these parameters, they are assigned uniform priors.
% Nothing is presumed known a priori about these parameters. Consequently, 
% improper uniform priors are assigned to the total number of signal events and 
% to the number of QCD background events in the $q\bar{q}^{\prime}\,b\bar{b}$ channel.  
The posterior density is then integrated over all parameters except 
$\sigma_{V\!H^{0}}^{\prime}$, and a 95\% credibility level (C.L.) upper limit on 
$\sigma_{V\!H^{0}}^{\prime}$ is obtained by calculating
the 95$^{\rm th}$ percentile of the resulting distribution.  When 
combining channels, the same procedure is applied to the product of their
likelihoods.  Correlations in the total efficiencies are taken
into account by identifying common parameters such as the $b$-tagging efficiency
and some kinematical efficiencies.  Each of these common parameters is then 
assigned a single prior.

Upper limits on $\sigma_{V\!H^{0}}\times\beta$ in each channel and in all channels
combined are summarized in Table \ref{table:vhsum} as a function of $H^0$
mass.  
\begin{table}[htb!]
\caption{\label{table:vhsum}
The 95\% credibility level upper limits on 
$\sigma(p\bar{p}\rightarrow VH^{0})\times\beta$ where 
$\beta = {\rm BR}(H^{0}\rightarrow b\bar{b})$, for each of the search channels 
and their combination, as a function of $H^0$ mass, $M_{H}$ (GeV/$c^{2}$).  
Also shown are the expected limits under the assumption of no $H^0$ signal. ST
designates the single $b$-tagged subsample and DT the double $b$-tagged subsample.
}
\begin{ruledtabular}
\begin{tabular}{l!{\extracolsep{3mm minus 3mm}}|D{,}{}{-1}D{,}{}{-1}D{,}{}{-1}}
   & \multicolumn{3}{c}{Measured (expected) upper limits (pb)} \\ 
Channel                          & M_H\; ,=\; 90   & M_H\; ,=\; 110  & M_H\; ,=\; 130  \\
\hline
$\ell^{+}\ell^{-}\,b\bar{b}$     & 55.6,\;\; (36)  & 31.8,\;\; (24)  & 23.8,\;\; (25)  \\
$\nu\bar{\nu}\,b\bar{b}$ (ST)    & 20.8,\;\; (30)  & 20.8,\;\; (21)  & 18.4,\;\; (17)  \\
$\nu\bar{\nu}\,b\bar{b}$ (DT)    & 10.4,\;\; (17)  &  9.2,\;\; (14)  &  8.0,\;\; (12)  \\ 
$\nu\bar{\nu}\,b\bar{b}$ (ST+DT) &  7.6,\;\; (13)  &  7.8,\;\; (11)  &  7.4,\;\; (8.8) \\ 
$\ell\nu\,b\bar{b}$ (ST)         & 30.0,\;\; (18)  & 29.4,\;\; (15)  & 27.6,\;\; (12)  \\
$\ell\nu\,b\bar{b}$ (DT)         & 31.0,\;\; (24)  & 26.6,\;\; (19)  & 24.2,\;\; (18)  \\ 
$\ell\nu\,b\bar{b}$ (ST+DT)      & 23.2,\;\; (13)  & 22.6,\;\; (11)  & 21.6,\;\; (9.0) \\ 
$q\bar{q}^{\prime}\,b\bar{b}$    & 38.2,\;\; (77)  & 21.2,\;\; (43)  & 17.8,\;\; (29)  \\
\hline
All combined                     &  7.8,\;\; (7.1) &  7.2,\;\; (5.7) &  6.6,\;\; (4.7) \\
\end{tabular}
\end{ruledtabular}
\end{table}

\begin{figure}[htb!]
\includegraphics[width=0.9\columnwidth]{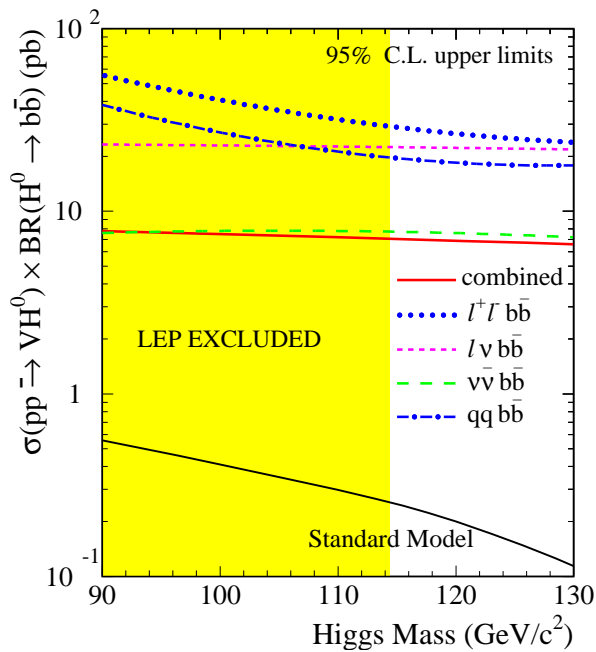}
\caption{\label{fig:vh_fig3}
Summary of all Run I CDF $95\%$ credibility level upper limits on 
$\sigma(p\bar{p} \rightarrow VH^0)\cdot \beta$.  The lines for the 
$\nu\bar{\nu}\,b\bar{b}$ and $\ell\nu\,b\bar{b}$ channels represent 
the combined limits from the single $b$-tagged and double $b$-tagged
subsamples.  Shown for comparison is the standard model prediction,
and the region excluded by the LEP experiments. }
\end{figure}

These results are also plotted in Figure \ref{fig:vh_fig3}.  The
standard model prediction is about 30 times smaller than the measured
95\% C.L. upper limits.  For the $\ell\nu\,b\bar{b}$ and 
$q\bar{q}^{\prime}\,b\bar{b}$ channels, 
the limits reported here are slightly different from
those previously published \cite{vhlnbb,vhqqbb};  this is mainly due to
our improved understanding of the $b$-tagging efficiency \cite{topxs2}.
Table \ref{table:vhsum} also shows expected upper limits under the 
assumption of zero signal.  
%These expectations are based on prior information
%only, and can therefore not be computed for the $q\bar{q}^{\prime}b\bar{b}$
%channel since the mistag background in the latter is a priori unknown.
These expectations are calculated over an ensemble of
experiments similar to this one, but where the background normalizations
are fluctuated around their expected values by their uncertainties.
%Finally, to understand the sensitivity of the cross section limits to our
%background estimates we note that by increasing the backgrounds in
%all channels except $q\bar{q}^{\prime}\,b\bar{b}$ by 20\% the combined limits
%decrease by about 20\%.
We note that the observed combined limits are driven by the $\nu\bar{\nu}\,b\bar{b}$
channel, as a result of an observed downward fluctuation in
the dijet invariant mass region of interest in this channel.
However, the expectation is for the $\nu\bar{\nu}\,b\bar{b}$ and
$\ell\nu\,b\bar{b}$ channels to have comparable sensitivity.
%\begin{figure}[htb!]
%\includegraphics[width=0.9\columnwidth]{vh_limits_prl_v3.eps}
%\caption{\label{fig:vh_fig3}
%Summary of all Run I CDF $95\%$ credibility level upper limits on 
%$\sigma(p\bar{p} \rightarrow VH^0)\cdot \beta$.  The lines for the 
%$\nu\bar{\nu}\,b\bar{b}$ and $\ell\nu\,b\bar{b}$ channels represent 
%the combined limits from the single $b$-tagged and double $b$-tagged
%subsamples.  Shown for comparison is the standard model prediction,
%and the region excluded by the LEP experiments. }
%\end{figure}

%conclusion
%----------
In conclusion, we have searched for $Z^0 H^0$ production using the
$\ell^+ \ell^-$ and $\nu\bar{\nu}$ decay channels of the $Z^0$ and
produced limits on $VH^0$ production using these channels.
We combined these limits with those previously published using other 
decay channels of the vector bosons to obtain final CDF Run I 
95\% C.L. limits on $\sigma_{V\!H^{0}} \times \beta$ ranging from 7.8\,pb
to 6.6\,pb for $H^0$ masses of $90\,{\rm GeV}/c^2$ to $130\,{\rm GeV}/c^2$.
These limits additionally apply to any scalar particle decaying
to $b\bar{b}$ that is produced in association with a vector boson.
These results and the combination methodology establish the 
foundation for our searches in the Tevatron Run\,II data at $\sqrt{s} = 1.96$\,TeV 
which are exploiting more search channels, an improved detector, and more
advanced analysis techniques~\cite{higgsstudy}.

We thank the Fermilab staff and the technical staffs of the
participating institutions for their contributions.  This work was
supported by the U.S. Department of Energy and National Science Foundation;
the Italian Istituto Nazionale di Fisica Nucleare; the Ministry of Science,
Culture, and Education of Japan; the Natural Sciences and Engineering Research
Council of Canada; the National Science Council of the Republic of China;
and the A. P. Sloan Foundation.

%-----------------------
%  Bibliography
%-----------------------


\begin{thebibliography}{99}
\bibitem{eta} In the CDF coordinate system, $\theta$ and $\phi$ are the 
polar and azimuthal angles, respectively, with respect to the proton beam 
direction ($z$ axis). The pseudorapidity $\eta$ is defined as 
$-\ln \tan(\theta/2)$.  The transverse momentum of a particle is 
$\Pt = P \sin\theta$.  The analogous quantity using calorimeter energies, 
defined as $\Et = E \sin\theta$, is called transverse energy.  The missing 
transverse energy, $\met$, is defined as $-\sum \Et^i \,\hat{n}_i$, where 
$\hat{n}_i$ is the unit vector in the transverse plane pointing from the 
interaction point to the energy deposition in calorimeter cell $i$.
\end{thebibliography}
\end{document}